\documentclass[12pt]{article}

\usepackage{graphicx,psfrag,color}       
\usepackage{multirow,rotating}

\def\-{~-~}
\def\+{~+~}
\def\={~=~}
\def\eq{~\equiv~}
\def\beq{\begin{equation}}
\def\eeq{\end{equation}}
\def\beqn{\begin{eqnarray}}
\def\eeqn{\end{eqnarray}}
\def\bb{\begin{eqnarray*}}
\def\ee{\end{eqnarray*}}
\newcommand{\calle}[1]{(\ref{#1})}
\newcommand{\lp}{\left(}
\newcommand{\lb}{\left\lbrack}

\newcommand{\rp}{\right)}
\newcommand{\rb}{\right\rbrack}

\newcommand{\CR}{\mathcal{R}}

\newcommand{\CG}{\mathcal{G}}

\begin{document}

\begin{center}
{\Large\bf ADDITION LAWS IN INTRODUCTORY PHYSICS}

\vspace{5mm}
     
{\bfseries\large C.J. Efthimiou}\footnote{costas@physics.ucf.edu}
 ~and~
{\bfseries\large R.A. Llewellyn}\footnote{ral@physics.ucf.edu}\\
     Department of Physics\\
     University of Central Florida\\
     Orlando, FL 32816
\end{center}

\vspace{2cm}

\begin{abstract}
  We propose a unified approach to addition of 
  some physical quantities (among which resistors and capacitors are
  the most well-known) that 
  are usually encountered in introductory physics
  such that the formul\ae\ required to solve problems
  are always simply additive. 
  This approach has the advantage of being consistent with the intuition
  of students. 
  To demonstrate the effectiveness of our approach,
  we propose and solve several problems. 
  We hope that this article can serve as a resource paper for problems
  on the subject.
\end{abstract}

\section{Introduction}

All introductory physics textbooks, with or without calculus, cover the addition
of both resistances and capacitances in series and in parallel.
The formul\ae\ for adding resistances
\beqn
\label{eq:1}
   R &=& R_1 + R_2 + \dots~, 
                          ~~~~~~~~~\mbox{in~series}~, \\
\label{eq:2}
  {1\over R} &=& {1\over R_1} + {1\over R_2} + \dots~, 
                          ~~~~~~~~\mbox{in~parallel}~,
\eeqn
and capacitances
\beqn
\label{eq:3}
  {1\over C} &=& {1\over C_1} + {1\over C_2} + \dots~,
                         ~~~~~~~~\mbox{in~series}~,\\
\label{eq:4}
   C &=& C_1 + C_2 + \dots~, 
                          ~~~~~~~~~\mbox{in~parallel}~,
\eeqn
are well-known and well-studied in all the books.

In books with calculus there are often end-of-chapter problems in which
students must
find $R$ and $C$  using continuous versions of equations 
\calle{eq:1} and \calle{eq:4} \cite{HRW,Serway,Tipler,WP,YF}. However, 
we have found \textit{none}  which includes problems that make
use of continuous versions of equations \calle{eq:2} and \calle{eq:3}
\cite{HRW,Hecht,Nolan,Serway,Tipler,WP,YF}.
Students who can understand and solve the first class of problems 
should be able
to handle the second class of problems, as well.
We feel that continuous problems that make use of all four equations
should be shown to the students in order to give them a 
global picture of how calculus is applied to physical problems.
Physics contains much more than mathematics. When integrating quantities
in physics,
\textit{the way we integrate them is motivated by the underlying physics}.
Students often forget
the physical reasoning and they tend to add  (integrate) quantities
only in one way.

In this paper, we introduce an approach to solving continuous
versions of equations \calle{eq:2} and \calle{eq:3} that is as straightforward
and logical
for the students as solving continuous versions of equations \calle{eq:1} 
and \calle{eq:4}. We then extend the logic to the addition of other
quantities encountered in undergraduate introductory physics. This 
demonstrates that the method is not specific to resistors and capasitors but 
general and includes all quantities obeying similar addition laws.

The organization of this article is as follows: section \ref{sec:formulae}
discusses many  physical quantities taken from introductory physics
that obey similar addition laws. Among these quantities, resistance and 
capacitance are the most well-known to students. Inductance is also known
but probably not mastered at the level it should be. Elasticity is somewhat
known, but thermal resistance,
diffusion resistance, and viscous resistance are almost unkown to
students. As a result, for resistance and capacitance we only remind
readers of the basic formul\ae, while for the rest quantities we expand the discussion
to some length so the students will become familiar with the physical 
background. In section \ref{sec:examples}, we present basic applications
of the addition formul\ae. This section is meant to demonstrate
in a simple way how one chooses the correct addition
formula (in series or in parallel), given a problem.
In section \ref{sec:problems} we solve several additional problems that
make use of the main idea. 
In each problem, we have chosen one representative quantity. 
However, the reader must realise that the same problem can be stated for any 
of the quantities given in section \ref{sec:formulae}, not just the chosen 
quantity. 

We  hope 
that this article will motivate teachers to explain to students the subtle
points between `straight integration' as taught in calculus and 
`physical integration' to find a 
physical quantity.

\section{Basic Formul\ae}
\label{sec:formulae}

\subsection{Resistance}

The basic formula to compute resistance is the formula of a uniform cylindrical
resistor:
$$
   \fbox{$\displaystyle
   R\=\rho\, {L\over A}
         $}~,
$$
where $\rho$ is the resistivity of the material, $L$ is the length of the
conductor and $A$ the cross-sectional area. 
Written as conductance\footnote{Often the term \textit{conductivity}
is used for $G$. However, the term \textit{conductance} is in uniform
lingustic aggreement with the rest of the terminology.}, this is 
$$
     G \= {1\over R}~,
$$
is
$$
   \fbox{$\displaystyle
   G\= \sigma\, {A\over L}
         $}~,
$$
where $\sigma=1/\rho$ is the conductivity of the material.

\subsection{Capacitance}

The basic formula to compute  capacitance is that of a parallel-plate
capacitor filled with a uniform dielectric material:
$$
   \fbox{$\displaystyle
   C\=\varepsilon_0\,\kappa\, {A\over d}
         $}~,
$$
where $\kappa$ is the dielectric constant of the dielectric, $A$ the area
of the plates, and $d$ the distance between the plates.
We are going to call the inverse of the capacitance
$$
   D \= {1\over C}~,
$$
the {\bfseries incapacitance} of the capacitor. For a parallel-plate
capacitor
$$
   \fbox{$\displaystyle
   D\={1\over\varepsilon_0\,\kappa}\, {d\over A}
         $}~.
$$

\subsection{Inductance}

Calculation of inductance is usually based on the definition
\beq
    L\={\Phi_B\over I}~,
\label{eq:Ldef}
\eeq
and therefore some discussion is necessary regarding our point of view.

It is well-known that the inductance for a solenoid is given by
\beq
   \fbox{$\displaystyle
   L\= \mu_0\,N^2\, {A\over \ell}
         $}~,
\label{eq:Lsolenoid}
\eeq
where $N$ is the number of turns of the solenoid, $A$ is the cross-section
and $\ell$ the length
of the solenoid. 
This expression can be used as the basic formula
when we compute the inductance of another geometry which involves
an inductor made from a single wire by twisting it in a particular geometry
and creating a number $N$ of turns.  

\begin{figure}[htb!]
\begin{center}
\psfrag{I}{$I$}
\psfrag{d}{$d$}
\psfrag{w}{$w$}
\psfrag{B}{$B$}
\psfrag{L}{$\ell$}
\includegraphics[width=8cm]{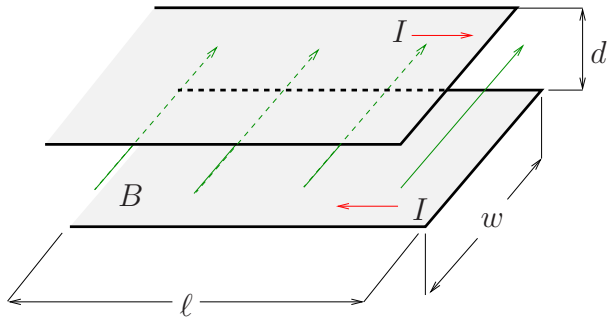}
\end{center}
\caption{A parallel-plate inductor.}
\label{fig:parallel-plates}
\end{figure}

When we have some geometry in which no obvious way to count `turns'  exists,
we must use a different formula. Clearly, the corresponding basic formula
must come
from a simple geometry giving rise to a uniform magnetic field.
In figure \ref{fig:parallel-plates} we see an inductor with such properties.
The inductor consists of two finite-plane wires that carry opposite currents
with uniform linear current density. 
Due to the complete analogy with a parallel-plate capacitor, we call
it the {\bfseries parallel-plate inductor}. 

An infinite sheet with uniform linear current density $J_s$ creates a uniform
magnetic field in space with value $B=\mu_0 J_s/2$. The direction of the field
is found using the right-hand rule. When two infinite sheets are given,
the total field is the sum of two fields. Given the directions of the
currents of the parallel-plate inductor, the magnetic field adds to zero outside the
plates and to $B=\mu_0 J_s$ between the plates. Of course, in the case
of finite plates this result is only approximate, being a good approximation
when $d\ll w$. To compute the inductance of the parallel-plate inductor,
we must compute the flux $\Phi_B$ in the definition \calle{eq:Ldef}.
We notice that the magnetic field is perpendicular to the area $A=\ell d$.
In this case
$$
   \Phi_B\=B\, A \= \mu_0 J_s \ell d
$$
and $I=J_s w$. Therefore
$$
   L\= {\Phi_B\over I}\=\mu_0 \,{\ell \,d\over w}~.
$$
Notice that the only difference is that
the `solenoid' is now made from only one turn, i.e. $N=1$. Therefore, in
problems where the current is distributed continuously only `one turn'
should be
counted. This issue can cause to curious students considerable frustration
since they can
raise the following objection: ``In the calculation of the inductance,
we concluded that 
$N=1$  since one `turn' was used (incorrectly) in the computation of flux,
namely only the flux through one cross-section was used.
However, there is an infinite number of cross-sections and an
integration has to
be done before
inserting the flux into equation \calle{eq:Ldef}.
Therefore, the result is not correct."
In this calculation, we believe that it is \textit{not} easy to
persuade the student
that she is wrong and no flux has been lost. And unfortunately,
similar situations
arise in all continuous problems. Books fall short in providing any
answers to this 
question---they just present the solution. Our method has the advantage,
among others, that  it answers this question
in a satisfactory and unambiguous way. It shows explicitly that all
flux has been counted. 
This  is easily understood if the reader
studies the problems of sections \ref{sec:parallel-ind} and
\ref{sec:coaxial}. Below, we present an independent calculation to confirm
that the result found above for the inductance of the parallel-plate inductor
is correct.

The alternative calculation will be based on the energy stored in the inductor.
The magnetic energy density for the parallel-plate inductor is 
$$
  u_B\={B^2\over2\mu_0}\={1\over2}\,\mu_0\,J_s^2~.
$$
The total energy stored in the magnetic field is then
$$
  U\=u_B\, w\,d\,\ell \={1\over2}\, \mu_0{\ell d\over w}\, I^2~.
$$
However, for any inductor, the energy stored in the magnetic field is also given by
$U=L I^2/2$. When this is compared to the previous result, we conclude that
\beq
   L\=\mu_0\,{\ell\,d\over w}~,
\label{eq:Lbasic}
\eeq
i.e. the same result as found above. 

We are going to call the inverse of the inductance
$$
   K \= {1\over L}~,
$$
the {\bfseries dedutance} of the inductor.
Therefore, the deductance of a parallel-plate inductor is
$$
   \fbox{$\displaystyle
   K\= {1\over\mu_0}\, {w\over \ell \, d}
         $}~.
$$

\subsection{Thermal Resistance}

Imagine a cylinder made from a uniform conducting material whose bases
are kept at different temperatures. Then, due to the temperature difference
$\Delta T$ between the bases, heat will flow from one base to the other. The
rate according to which heat is flowing, i.e.
$$
    I_{th}\= {\Delta Q\over \Delta t}~,
$$
is called the {\bfseries thermal current}. It is known, see for example 
\cite{Alonso,Tipler},
that
\beq
    I_{th}\=\sigma_{th}\, A \, {\Delta T\over L}~,
\label{eq:I1}
\eeq
where $L$ is the length of the cylinder, $A$ is its cross-section, and $\sigma_{th}$ is
a constant characteristic of the material, called the {\bfseries thermal
conductivity}. 
We define the inverse of the
thermal conductivity 
$\rho_{th}=1/\sigma_{th}$ to be the {\bfseries thermal resistivity} of the material.
Equation \calle{eq:I1} is sometimes referred to as 
{\bfseries Fourier's law} for the flow of energy.

\begin{figure}[htb!]
\begin{center}
\psfrag{A}{$A$}
\psfrag{l}{$L$}
\includegraphics[width=6cm]{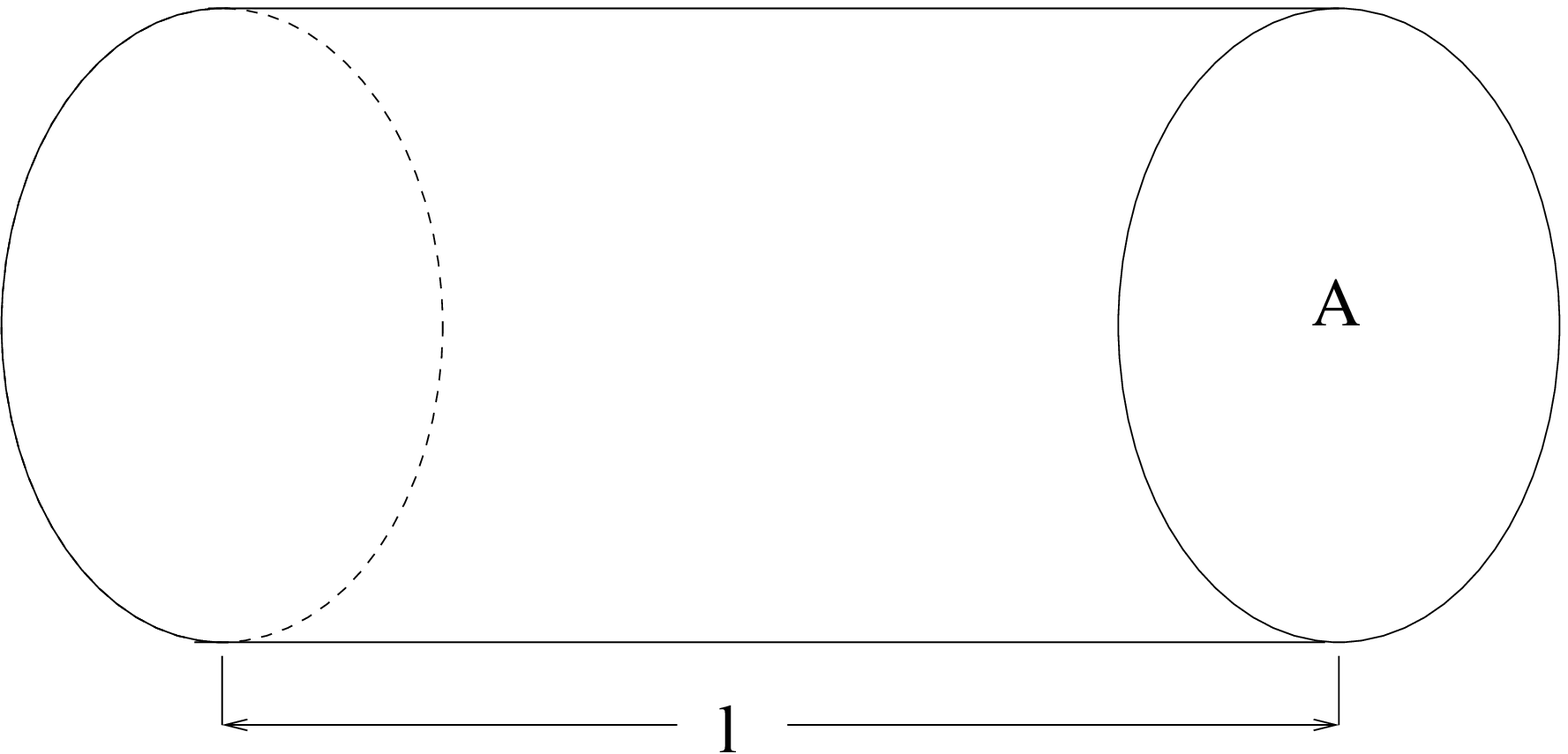}
\end{center}
\end{figure}

The {\bfseries thermal resistance} of the cylinder is then defined by
\beq
  \CR_{th} \= {\Delta T\over I_{th}}~.
\label{eq:I2}
\eeq
Notice the analogy with the standard resistance: $R=\Delta V/I$. 
Potential difference
is the reason electric current flows. Here, temperature difference is the
reason behind the thermal  current. Comparing the two formul\ae\ 
\calle{eq:I1} and \calle{eq:I2} we have written
above, we arrive at 
\beq
  \fbox{$\displaystyle
  \CR_{th} \=\rho_{th}\, {L\over A}
        $}~,
\label{eq:I7}
\eeq
an expression almost identical to that of the electric resistance for the cylinder.

We define the  {\bfseries thermal conductance} as
$$
    \CG_{th}\={1\over\CR_{th}}~.
$$
This implies that for the uniform cylinder
\beq
  \fbox{$\displaystyle
   \CG_{th}\=\sigma_{th}\, {A\over L}
        $}~.
\label{eq:I8}
\eeq

\subsection{Diffusion Resistance and Viscous Resistance}

The thermal conductivity  discussed in the last section is only a particular
example from a more general category of processes known as 
{\bfseries transport phenomena} \cite{Alonso}. Transport phenomena are 
irreversible processes that occur in systems that are not in statistical
equilibrium. In these systems, there is a net transfer of energy, matter, 
or momentum. 
Fourier's law stated in the previous section was a law for the flow of
energy. We will write similar laws for the flow of matter and the flow
of momentum. 

Imagine a cylinder filled with gas such that  the particle densities
$n_1$ and $n_2$ of the gas at the bases are kept constant at different
values. 
Then, due to the density difference
$\Delta n$ between the bases, particles will flow from one base to the other. The
rate according to which particles are flowing, i.e.
$$
    I_{diff}\= {\Delta n\over \Delta t}~,
$$
is called the {\bfseries particle current}. It is known, see for example
\cite{Alonso},
that
\beq
    I_{diff}\=\sigma_{diff}\, A \, {\Delta n\over L}~,
\label{eq:I3}
\eeq
where $L$ is the length of the cylinder, $A$ is its cross-section, and 
$\sigma_{diff}$ is
a constant characteristic of the material called the {\bfseries diffusion
coefficient}. Another name, in the spirit of  what we have been discussing, 
would be {\bfseries diffusion conductivity}. The inverse of $\sigma_{diff}$,
$\rho_{diff}=1/\sigma_{diff}$, is named the {\bfseries diffusion resistivity}
of the material.
Equation \calle{eq:I3} is sometimes referred to as
{\bfseries Fick's law}.
The {\bfseries diffusive resistance} of the cylinder is then defined by
\beq
  \CR_{diff} \= {\Delta n\over I_{diff}}~.
\label{eq:I4}
\eeq
Its inverse gives the  {\bfseries diffusive conductance}:
$$
    \CG_{diff}\={1\over\CR_{diff}}~.
$$

Now imagine that the thermal agitation (speed) of the molecules
at the two bases of the cylinder is different.
Then, due to the speed difference
$\Delta v$ between the bases, momentum will flow from one base to the other. The
rate according to which speed is flowing, i.e.
$$
    I_{vis}\= {\Delta v\over \Delta t}~,
$$
is called the {\bfseries momentum current}. It is known, see for example
\cite{Alonso},
that
\beq
    I\=\sigma_{vis}\, A \, {\Delta v\over L}~,
\label{eq:I5}
\eeq
where $L$ is the length of the cylinder, $A$ is its cross-section, 
and $\sigma_{vis}$ is
a constant characteristic of the material called the {\bfseries
coefficient of viscosity}. Another name, again in the spirit of  what we have been
discussing,
would be {\bfseries viscous conductivity}. The inverse of $\sigma_{vis}$,
$\rho_{vis}=1/\sigma_{vis}$, is the {\bfseries viscous resistivity} of the material.
The {\bfseries viscous resistance} of the cylinder is then defined by
\beq
  \CR_{vis} \= {\Delta v\over I}~.
\label{eq:I6}
\eeq
Its inverse gives the  {\bfseries viscous conductivity}:
$$
    \CG_{vis}\={1\over\CR_{vis}}~.
$$

Formul\ae\ \calle{eq:I7} and \calle{eq:I8} we derived in the previous section
are also applicable in the present cases with the appropriate index changes. 

Although it is not the topic of our article, we mention that ultimately all transport 
phenomena are related via the microscopic description. In particular we can find
the following expressions for the coefficients:
\bb
   \sigma_{th} &=& {1\over 2}\,n\,k_B\, v_{ave}\, l_{free}~,\\
   \sigma_{diff} &=& {1\over 3}\, v_{ave}\, l_{free}~,\\
   \sigma_{vis} &=& {1\over 3}\,n\,m\, v_{ave}\, l_{free}~,
\ee
where $l_{free}$ is the mean free path of the molecules, $v_{ave}$ the average
velocity of the molecules, $n$ the number of molecules per unit volume,
$m$ the mass of one molecule, and $k_B$ Boltzmann's constant. These expressions
should be compared with that of the electric conductivity which is more
familiar:
$$
  \sigma \= {1\over 2}\,n\,{q^2\over m}\,\, {l_{free}\over v_{ave}}~,
$$
where $q$ is the charge of a carrier.

\subsection{Elasticity}

The concept of elasticity is more than a mere definition.
The behavior of a  rubber band or the behavior of a rod or a cable under stress is
basically analogous to that of many  springs connected together.

\begin{figure}[htb!]
\begin{center}
\psfrag{A}{$A$}
\psfrag{l}{$L$}
\psfrag{dx}{$dx$}
\psfrag{x}{$x$}
\includegraphics[width=6cm]{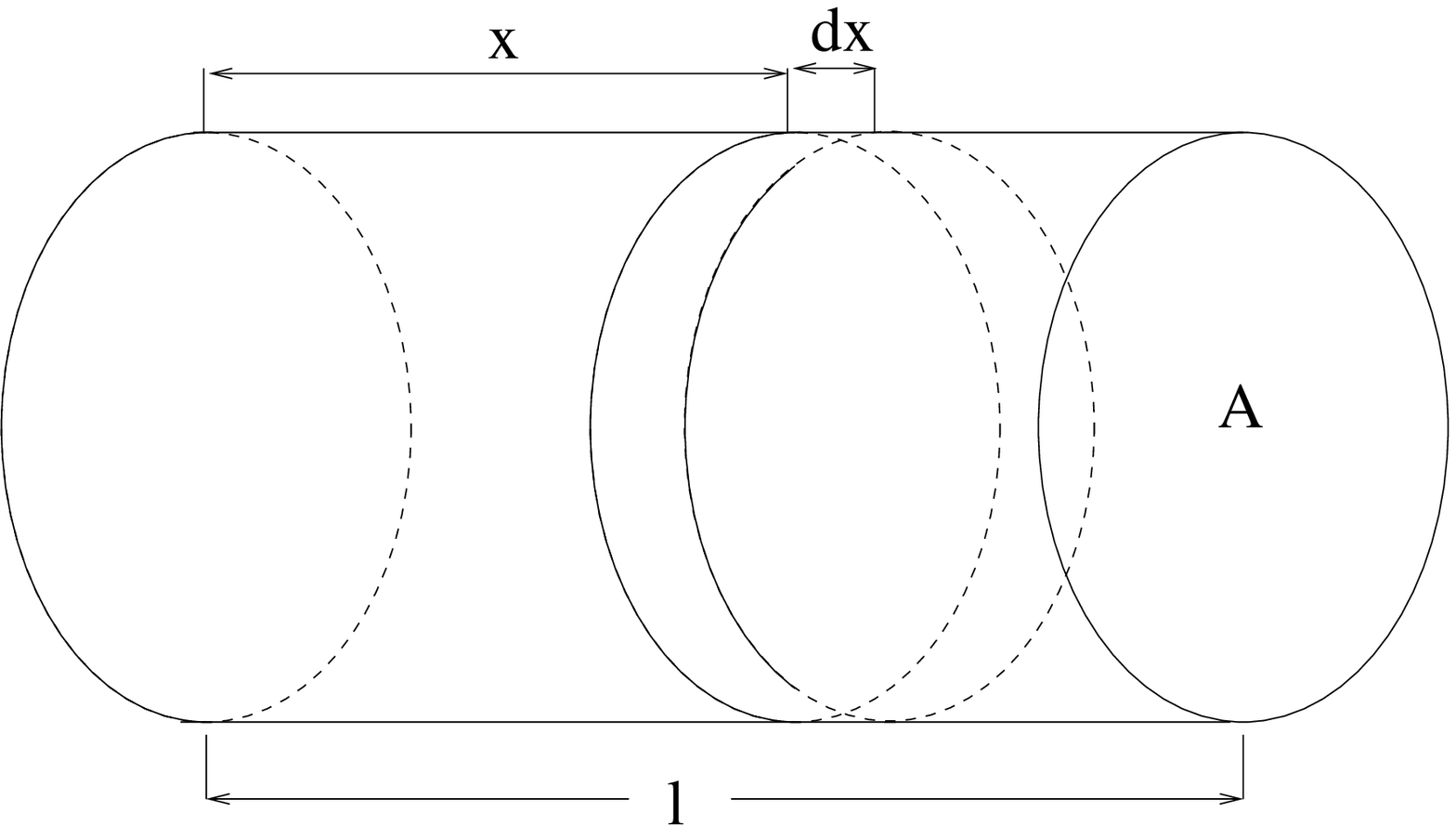}
\end{center}
\end{figure}

Let's imagine a uniform rod of length $L$ and cross-sectional area $A$. 
We focus on an infinitesimal piece of length
$dx$ at distance $x$ from one base.
If $d\xi$ is the infinitesimal extension of this piece under the force
$F(x)$, then Hooke's law states that
$$
    d\xi \= - F(x) \, d\ell~,
$$ 
where $d\ell$ is the elasticity constant for the piece $dx$. We can
write the above relation as
$$
    {d\xi\over dx} \= - F(x) \, {d\ell\over dx}~,
$$
where $\lambda=d\ell/dx$ is the elasticity per unit length and 
$\epsilon=d\xi/dx$ is the extension of the system per unit length, known
as {\bfseries linear strain}. 
It is known that approximately \cite{Alonso} 
$$
   {d\xi\over dx} \= - {1\over Y A}\, F(x)~,
$$
where $A$ is the cross-section and $Y$ is a constant characteristic of
the material known as {\bfseries Young's modulus}.
Combining the last expressions we conclude that
$$
   \lambda \= {d\ell\over dx} \= {1\over Y A} ~.
$$
If $\lambda$ is constant then
$$
    \fbox{$\displaystyle
    \ell\= {1\over Y}\, {L\over A}
          $} ~,
$$
where $L$ is the length of the system. The stiffness would be respectively:
$$ 
   \fbox{$\displaystyle
   k\= Y\, {A\over L}
         $} ~.
$$
We have thus obtained basic formul\ae\ similar to those of resistors and capacitors
that allow the computation of $k$ and $\ell$ in any geometry. Most probably
these formul\ae\ are well-known to engineers, but they are not well-known
among physicists. However, once written down, they look familiar  and
natural.

\section{Addition Formul\ae}
\label{sec:examples}

Let $R_1, R_2, ...$ be some resistors\footnote{By abuse of language,
                       we identify the 
                      objects with their property.},
    $C_1, C_2, ...$ be some capacitors,
    $L_1, L_2, ...$ be some inductors,
    $k_1, k_2, ...$ be some springs,
    $\CR_1,\CR_2, ...$ be some transport conductors (either thermal,
                      or diffusion, or viscous).
We will also indicate by $R$, $C$, $L$, $k$, $\CR$ the
equivalent resistance, capacitance, inductance, stiffness, and transport
resistance, respectively, either when the objects are connected
in series or in parallel. Either from introductory physics, or as a
straightforward exercise on the definitions, the reader can persuade himself
that when the objects are connected in series
\bb
   R &=& R_1 + R_2 + \dots~,\\
   {1\over C} &=& {1\over C_1} + {1\over C_2} + \dots~,\\
   L &=& L_1 + L_2 + \dots~,\\
   {1\over k} &=& {1\over k_1} + {1\over k_2} + \dots~,\\
   \CR &=& \CR_1 + \CR_2 + \dots~,
\ee
and when they are in parallel 
\bb
  {1\over R} &=& {1\over R_1} + {1\over R_2} + \dots~,\\
   C &=& C_1 + C_2 + \dots~,\\
  {1\over L} &=& {1\over L_1} + {1\over L_2} + \dots~,\\
   k &=& k_1 + k_2 + \dots~,\\
  {1\over \CR} &=& {1\over \CR_1} + {1\over \CR_2} + \dots~.
\ee
Introducing the conductance, incapacitance, deductance, elasticity,
and thermal conductance, we can rewrite them in a form that is
always additive:
$$
    P \= P_1 + P_2 + \cdots~,
$$
where,
\bb
  && \mbox{if~the~elements~are~connected~in~series},
     ~P~\mbox{stands~for~any~of}~
     R,D,L,\ell,\CR~,\\ 
  && \mbox{if~the~elements~are~connected~in~parallel},
     ~P~\mbox{stands~for~any~of}~
     G,C,K,k,\CG~. 
\ee

In the remaining section, we will demonstrate the application of the 
addition formul\ae\ by examining specific
examples from resistors and capacitors. Along with the solution,
several comments are made to help the reader understand some of the
implicit assumptions and other details in the solution.
In the section that follows, more problems are discussed for the readers who
wish to master the technique.

\newcommand{\paral}{{\bfseries\textcolor{red}{parallel}}}
\newcommand{\series}{{\bfseries\textcolor{red}{series}}}
\newcommand{\resistors}{{\bfseries\textcolor{blue}{resistors}}}
\newcommand{\capacitors}{{\bfseries\textcolor{blue}{capacitors}}}
\newcommand{\inductors}{{\bfseries\textcolor{blue}{inductors}}}
\newcommand{\springs}{{\bfseries\textcolor{blue}{springs}}}
\newcommand{\transport}{{\bfseries\textcolor{blue}{transport}}}
\newcommand{\thermal}{{\bfseries\textcolor{blue}{thermal}}}
\newcommand{\conductors}{{\bfseries\textcolor{blue}{conductors}}}

\begin{table}[p]
\begin{center}
\

\vspace{20mm}
\rotatebox{90}{
\vbox{
\begin{tabular}{|c|c|c|c|c|c|c|} \hline
         & & & & & & \\ 
         {\bfseries connection} & {}
    & \resistors & \capacitors & \inductors & \springs & \transport~\conductors \\ 
         & & & & & & \\ \hline
         & & & & & & \\ 
   \multirow{4}{14mm}[0mm]{\series} 
       & definition
       & $\matrix{ R={\Delta V\over I} \cr}$
       & $\matrix{ D={\Delta V\over Q}\cr}$
       & $\matrix{ L={\Phi_B\over I}\cr}$
       & $\matrix{ \ell={\xi\over F} \cr}$
       & $\matrix{ \CR_{tr}={\Delta T_{tr}\over I_{tr}}\cr}$  \\ 
       & & & & & & \\ 
       & & resistance & incapacitance & inductance
         & elasticity & thermal resistance  \\ 
       & & & & & & \\ 
       & basic formula  
       & $\matrix{ R=\rho\,{L\over A} \cr}$ 
       & $\matrix{ D={1\over\varepsilon_0\kappa}\,{d\over A}\cr}$
       & $\matrix{ L=\mu_0\,N^2\,{A\over L}\cr}$
       & $\matrix{ \ell={1\over Y}\,{L\over A} \cr}$ 
       & $\matrix{ \CR_{tr}=\rho_{tr}\,{L\over A}\cr}$ \\ 
        & & & & & & \\ \hline 
        & & & & & & \\ 
   \multirow{4}{14mm}[0mm]{\paral} 
      & definition
      & $\matrix{ G={I\over \Delta V}\cr}$
      & $\matrix{ C={Q\over \Delta V} \cr}$
      & $\matrix{ K={I\over\Phi_B}\cr}$
      & $\matrix{ k={F\over\xi} \cr}$
      & $\matrix{ \CG_{tr}={I_{tr}\over \Delta T_{tr}}\cr}$  \\
       & & & & & & \\ 
       & & conductance & capacitance & deductance
         & stiffness & thermal conductance  \\ 
       & & & & & & \\ 
      & basic formula
      & $\matrix{ G=\sigma\,{A\over L}\cr}$
      & $\matrix{ C=\varepsilon_0\kappa\,{A\over d} \cr}$
      & $\matrix{ K={1\over\mu_0 N^2}\,{L\over A}\cr}$
      & $\matrix{ k=Y\,{A\over L} \cr}$ 
      & $\matrix{ \CG_{tr}=\sigma_{tr}\,{A\over L}\cr}$ \\
      & & & & & & \\ \hline 
\end{tabular}
}
}
\end{center}
\caption{This table summarizes the additive physical quantities in the most
         common cases encountered in introductory physics. The quantities 
         that are not usually defined in the introductory books are
         the conductivity $G=1/R$, the incapacitance $D=1/C$, the deductance
         $K=1/L$, the elasticity constant $\ell=1/k$, and the the thermal
         conductivity
         $\CG_{tr}=1/\CR_{tr}$. The index $tr$ stands for \textit{transport}
         and it should be interpreted as a generic name for any of the
         three cases
         of thermal conductivity, diffusion, or viscosity.}
\label{table:2}
\end{table}

\begin{enumerate}

\item
{\bfseries Problem} [Cylindrical Resistor]\\
The cylindrical resistor shown in figure \ref{fig:1} is made such that the
resistivity
$\rho$ is a function of the distance $r$ from the axis.
What is the total resistance $R$ of the resistor?

\begin{figure}[h]
\begin{center}
\psfrag{r}{$r$}
\psfrag{dr}{$dr$}
\psfrag{a}{$a$}
\psfrag{l}{$l$}
\includegraphics[width=10cm]{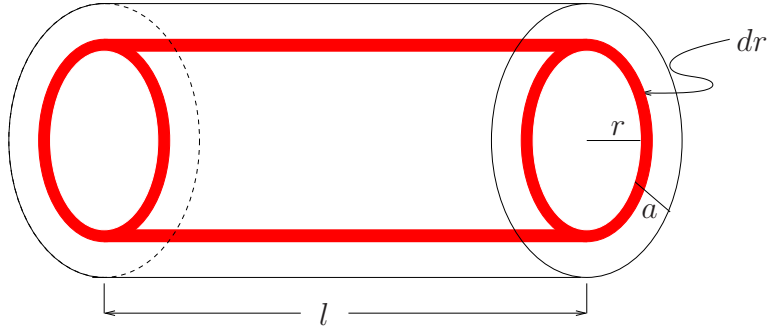}
\end{center}
\caption{The figure shows a  cylindrical wire of radius $a$. A potential 
         difference is
         applied between the bases of the cylinder and therefore electric
         current is running parallel to the axis of the cylinder.}
\label{fig:1}
\end{figure}

{\bfseries Solution}

We divide the cylindrical resistor into infinitesimal resistors in the
form of cylindrical shells of thickness $dr$. One of these shells is seen
in red  in figure \ref{fig:1}.
    
When the current is flowing along the axis of the cylinder,
the infinitesimal resistors are \textit{not} connected in series.
Instead, 
all of the infinitesimal cylindrical shells
of thickness $dr$  are connected at the same end points and, therefore,
have the same applied potential. In other words, the shells
are connected in parallel and it is the conductance that
is important. Specifically
$$
   G \=  \int_{\rm cylinder} dG~.
$$
For the infinitesimal shell
$$
  dG\=\sigma(r) \, {2\pi r dr\over\ell}~.
$$
Therefore
\bb
  G \=\int_{cylinder} dG \=  {2\pi\over\ell}\, \int_0^a  \sigma(r)\, r dr~.
\ee
For example, if $\sigma(r)\=\sigma_0 \,{a\over r}$, then
\bb
  G \= 2\sigma_0 \, {\pi a^2\over\ell} ~,
\ee
where $\sigma_0=1/\rho_0$.
The resistance is therefore 
\bb
    R\= {\rho_0\over2} \, {\ell\over \pi a^2}~.
\ee

\item
{\bfseries Problem} [Truncated-Cone Resistor]\\
A resistor is made from a truncated cone of material with uniform
resistivity 
$\rho$. 
What is the total resistance $R$ of the resistor when the potential 
difference is applied between the two bases of the cone?

{\bfseries Solution}

This is a well-known problem found in many of the introductory
physics textbooks \cite{HRW,Serway,Tipler,WP,YF}. 
We can partition the cone into infitesimal cylindrical resistors of
length $dz$. One representative resistor at distance $z$ from the top base
is seen in  figure \ref{fig:cone}. 
The area of the resistor is $A=\pi r^2$ and therefore its 
infinitesimal resistance is given by 
$$
   dR \= \rho\, {dz\over \pi r^2}~.
$$
From the figure we can see that
$$
  {z\over h}\= {r-b\over c-b} ~\Rightarrow~  dz\= {h\over c-b} dr ~.
$$

\begin{figure}[h]
\begin{center}
\psfrag{h}{$h$}
\psfrag{b}{$b$}
\psfrag{c}{$c$}
\psfrag{z}{$z$}
\psfrag{dz}{$dz$}
\psfrag{r}{$r$}
\includegraphics[width=7cm]{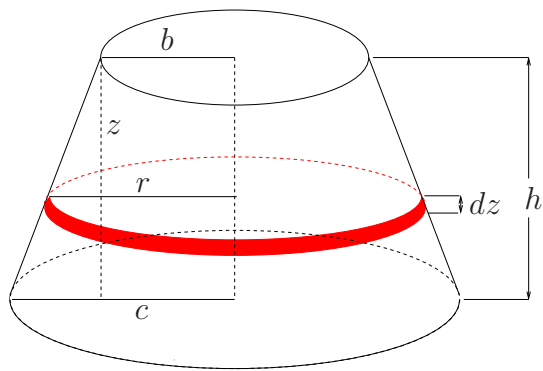}
\end{center}
\caption{A truncated cone which has been sliced in infitesimal cylinders of 
         height $dz$.}
\label{fig:cone}
\end{figure}

The infinitesimal resistors are connected in series and therefore
\beq
   R\=\int_{\rm cone} dR \= \rho\, {h\over\pi (c-b)} \, \int_b^c {dr\over r^2}
    \= \rho\,{h\over\pi bc} ~.
\label{eq:Rcone}
\eeq

{\bfseries Comment}:
{\footnotesize However, this solution, which is common in textbooks 
\cite{HRW,Serway,Tipler,WP,YF}, tacitly assumes that
the disks used in the partition of the truncated cone are equipotential
surfaces. This is of course not true, as can be seen quite easily.
If they were  equipotential surfaces, then the electric field lines
would be straight lines, parallel to the axis of the cone. However,
this cannot be the case as, close to the lateral surface of the cone,
it would mean that the current goes through the 
lateral surface and does not remain inside the resistor. Therefore,
the disks are not equipotential surfaces. One way out of this 
subtlety is to assume that
the disks are approximate equipotential surfaces as suggested in 
\cite{WP}. This is the attitude we adopt in this article as
our intention is not to discuss the validity of the partitions
used in each problem, but to emphasize the unified description
of resistances and capacitances as additive quantities. Similar
questions can be raised and studied in the majority of the problems 
mentioned in the present manuscript. A reader with serious interests in
electricity is referred to the article of  of Romano and Price
\cite{Romano} where the conical resistor is studied. Once that
article is understood, the reader can attempt to generalize
it to the rest of the problems of our article.}

\item
{\bfseries Problem} [Cylindrical Capacitor]

In introductory physics, the capacitance of a cylindrical capacitor
is found using the definition $C=Q/\Delta V$, where $Q$ is the charge 
on the positive plate of the capacitor and $\Delta V$ the absolute value of the
potential difference between the two plates. However, this problem
asks to compute the capacitance using only the formul\ae\ giving
the capacitance of a parallel plate, plane capacitor.

{\bfseries Solution}

As shown in the left side of figure \ref{fig:hollowcylinder},
the capacitor is partitioned into small
cylindrical capacitors for which
the distance between the plates is $dr$. For such small capacitors,
the formula of a parallel-plate capacitor is valid. We notice though that
all infintesimal capacitors are connected in series. Therefore
$$
    dD ~=~ {1\over\varepsilon_0}\, {dr\over 2\pi r h}~.
$$
and
$$
    D ~=~ \int_{\rm cylinder} dD
      ~=~ {1\over2\pi\varepsilon_0h}\, \int_a^b{dr\over r}
      ~=~ {1\over2\pi\varepsilon_0h}\, \ln{a\over b}~.
$$
The total capacitance is then
$$
    C ~=~ {1\over D} ~=~ {2\pi\varepsilon_0 h\over\ln{b\over a}}~.
$$

\begin{figure}[htb!]
\begin{center}
\psfrag{r}{$r$}
\psfrag{dr}{$dr$}
\psfrag{a}{$a$}
\psfrag{b}{$b$}
\psfrag{h}{$h$}
\includegraphics[height=10cm]{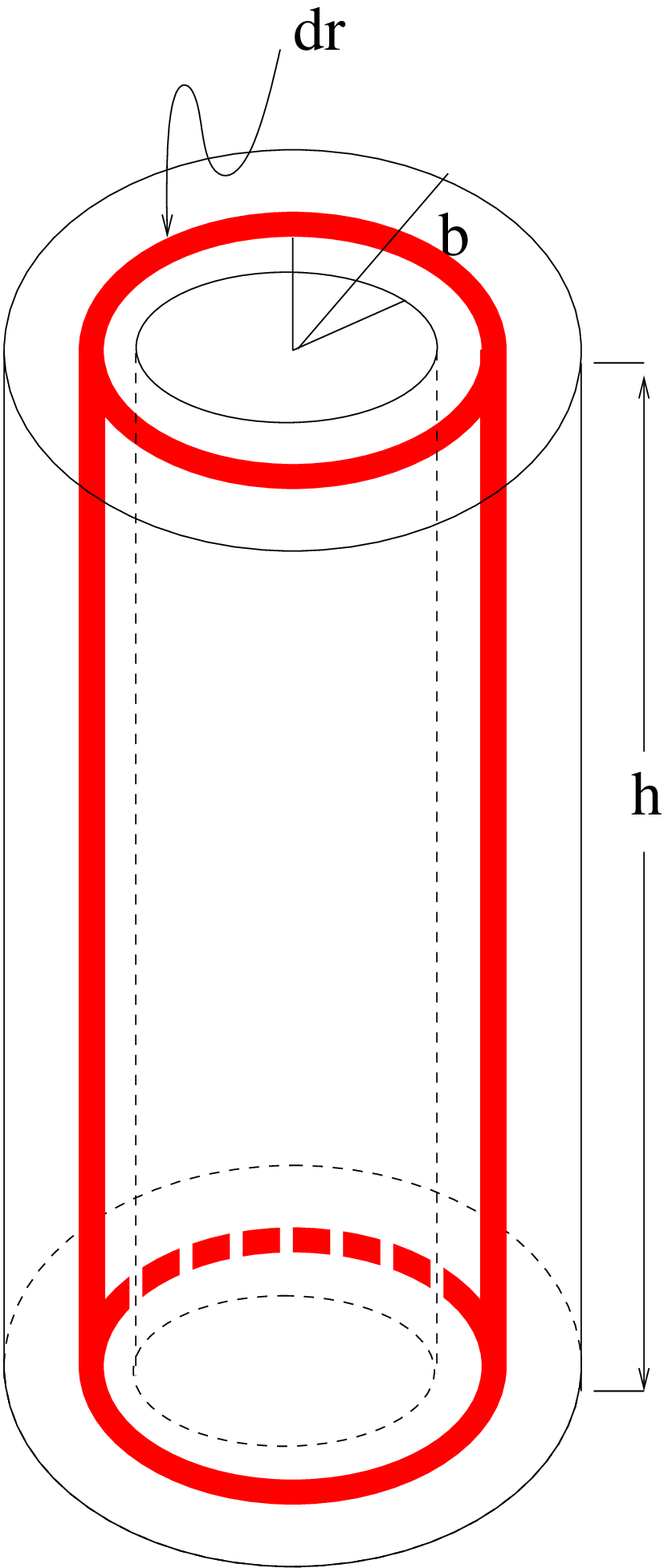}
\hspace{20mm}
\psfrag{r}{$r$}
\psfrag{dr}{$dr$}
\psfrag{a}{$a$}
\psfrag{b}{$b$}
\psfrag{h}{$h$}
\psfrag{dz}{$dz$}
\includegraphics[height=9cm]{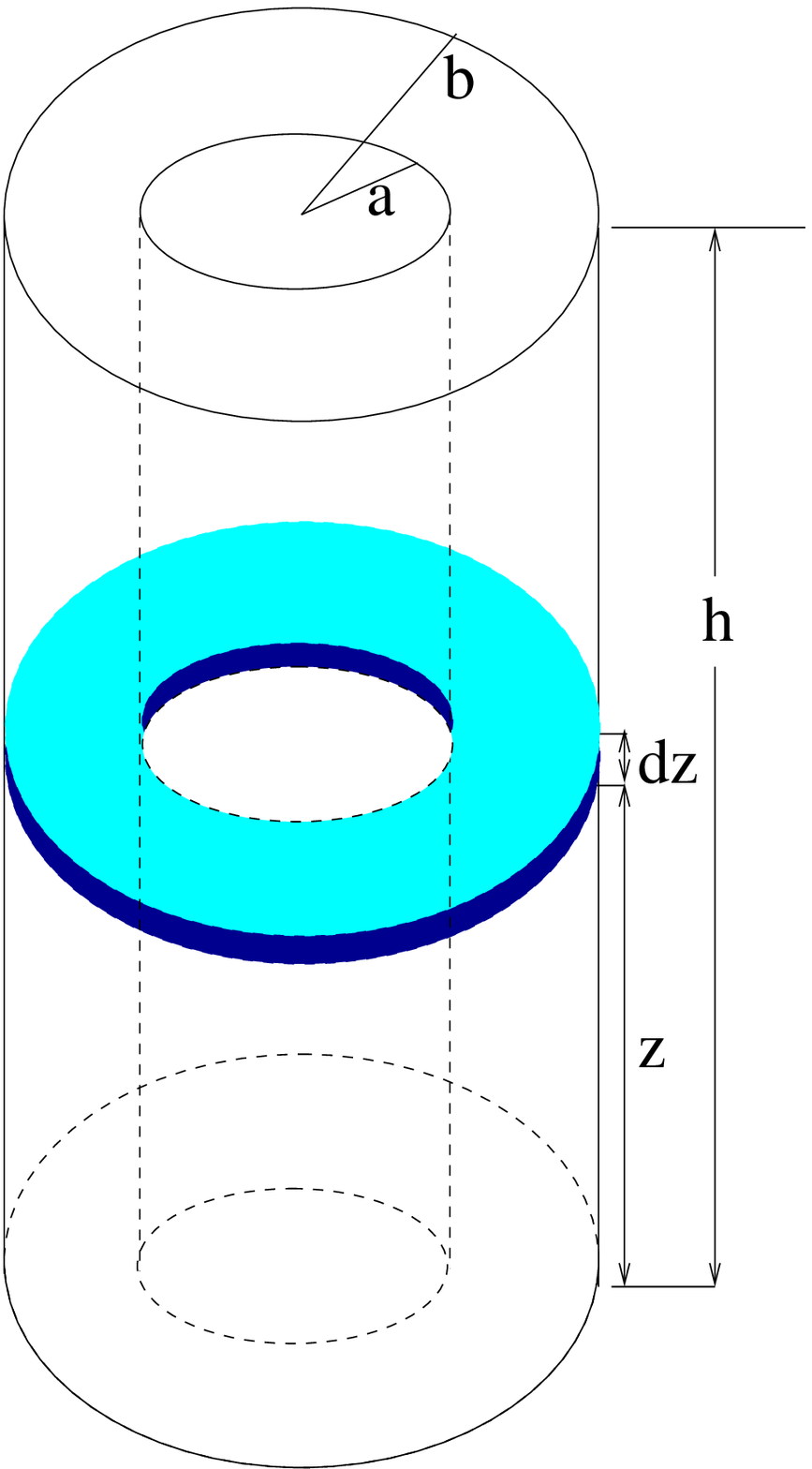}
\end{center}
\caption{A cylindrical capacitor with radii $a$ and $b$ and height $h$.
         In the left picture, we have sliced it in infinitesimal cylindrical
         shells, while in the right picture we have sliced it in infinitesimal
         annuli.}
\label{fig:hollowcylinder}
\end{figure}

{\bfseries Comment}:
\footnotesize
One might be tempted to
partition the cylindrical capacitor into infinitesimal
capacitors 
as seen in the figure to the right (blue section). Such capacitors
look simpler than the  
infitesimal cylindrical shell we used above. Furthermore, they
are connected in
parallel (notice that each capacitor is carrying an infinitesimal charge $dQ$
and $\int_{\rm cylinder} dQ=Q$) and  therefore it is enough to deal with
capacitance,
$C=\int_{\rm cylinder} dC$, and not incapacitance $D$.

However, with a minute's reflection the reader will
see that in order to use the parallel-plate capacitor formula in the
infinitesimal case, the distance between the plates must be infinitesimal
which indicates that the infinitesimal capacitors must
be connected in series.
In the proposed (blue) slicing, the distance between the plates
of the infinitesimal capacitor is finite, namely $b-a$. The infinitesimal
capacitor is still a cylindrical capacitor of infitesimal height and therefore
its capacitance should be expressed in a form that is  not known before the
problem is solved. In other words,
\beq
   dC ~=~ {2\pi\varepsilon_0\over\ln(b/a)} \, dz
\label{eq:unknown}
\eeq
from which
\beq
   C ~=~ {2\pi\varepsilon_0\over\ln(b/a)} \, \int_0^h dz
     ~=~ {2\pi\varepsilon_0\over\ln(b/a)} \,h~.
\label{eq:known}
\eeq
However, the expression \calle{eq:unknown} is unkown until the
result \calle{eq:known} is found.
\normalsize

\item
{\bfseries Problem} [Truncated-cone Capacitor]

A capacitor is made of two circular disks of radii $b$ and $c$ respectively
placed at a distance $h$ such that the line that joins their centers is
perpendicular to the disks. Find the capacitance of this arrangement
(Seen in figure \ref{fig:cone}).

{\bfseries Solution}

We partition the capacitor into infinitesimal
parallel-plate capacitors of distance $dz$ and plate area $A=\pi r^2$
exactly as seen in figure \ref{fig:cone}. These infitesimal
capacitors are connected in series and therefore the incapacitance
is the relevant additive quantity:
$$
   dD \= {1\over\varepsilon_0}\,{dz\over \pi r^2}~.
$$
Notice that the computation is identical to that of $R$ with final result:
\beq
  D \= {1\over\varepsilon_0}\, {h\over\pi bc}
  ~\Rightarrow~ C\= \varepsilon_0 \, {\pi bc\over h}~.
\label{eq:trunccone}
\eeq
When $b=c$, we recover the result of the parallel-plate capacitor.
\end{enumerate}

\section{Problems with Solutions}
\label{sec:problems}

In this section we pose and solve a number of problems that
will help the reader become fluid in the application of the  
quantities in table \ref{table:2}. A few of the problems are 
well-known, standard ones found in all textbooks;
these problems
are re-examined and solved in this section using our technique.

\subsection{Spherical Capacitor}

Re-derive the well-known expression for the capacitance of a spherical capacitor
$$
    C \= 4\pi\varepsilon_0 {ab\over b-a}~,
$$
(where $a, b$ are the radii of the spheres with $b>a$)
by partitioning it into
infinitesimal capacitors.

\footnotesize {\bfseries Solution} 

We partition the capacitor into spherical shells of thickness $dr$. The infinitesimal
shells are connected in series and therefore their 
incapacitance  is
$$
    dD \= {1\over\varepsilon_0}\,{dr\over 4\pi r^2}~.
$$ 
The total incapacitance is thus
\bb
    D \= \int_{\mbox{sphere}} dD \= {1\over4\pi\varepsilon_0}\, \int_{\mbox{sphere}}
         {dr\over  r^2} \= {1\over4\pi\varepsilon_0}\, {b-a\over ab}~,
\ee 
from which we find the well-known formula for the capacitance:
$$
   C \= 4\pi\varepsilon_0 {ab\over b-a}~.
$$
\normalsize

\begin{figure}[h]
\begin{center}
\psfrag{r}{$r$}
\psfrag{dr}{$dr$}
\psfrag{a}{$a$}
\psfrag{b}{$b$}
\includegraphics[width=7cm]{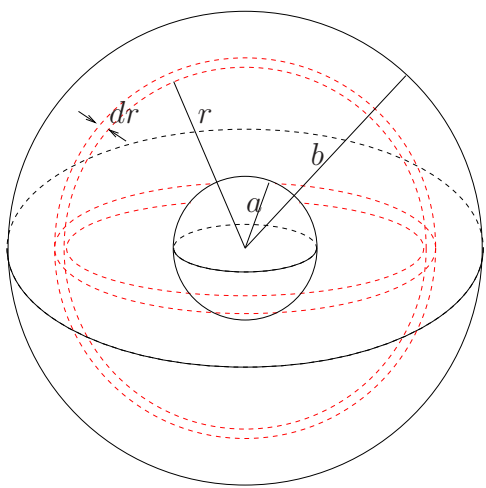}
\end{center}
\caption{The partition for a spherical capacitor into infinitesimal shells.}
\label{fig:spheres}
\end{figure}

\subsection{Spherical Capacitor with Dielectric}

Show that the capacitance of a spherical capacitor which is filled with
a dielectric having dielectric constant $\kappa(r)=c r^n$, where $r$
is the distance from the center and $c, n$ are constants, is given
by
$$
    C \= 4\pi\varepsilon_0 \, {c\over\ln(b/a)}
$$
for $n=-1$ and
$$
    C \= 4\pi\varepsilon_0 \, c(n+1)\, 
         {a^{n+1} b^{n+1}\over b^{n+1}-a^{n+1} }
$$
for $n\ne-1$.

\footnotesize {\bfseries Solution} 

As in the previous problem, we partition the capacitor into spherical shells of
thickness $dr$. The infinitesimal
shells are connected in series and therefore their 
incapacitance  is
$$
    dD \= {1\over\varepsilon_0\kappa(r)}\,{dr\over 4\pi r^2}~.
$$ 
The total incapacitance is thus
\bb
    D \= \int_{\mbox{sphere}} dD \= {1\over4\pi\varepsilon_0}\, \int_{\mbox{sphere}}
         {dr\over \kappa(r) r^2} \= 
      \= {1\over4\pi\varepsilon_0 c}\, \int_a^b {dr\over r^{n+2}}~.
\ee 
For $n=-1$,
$$
   D\={1\over4\pi\varepsilon_0 c}\, \int_a^b {dr\over r}
    \={1\over4\pi\varepsilon_0 c}\, \ln{b\over a}~.
$$
For $n\ne-1$,
\bb
    D \={1\over4\pi\varepsilon_0 c}\, \int_a^b {dr\over r^{n+2}}
      \={1\over4\pi\varepsilon_0 c}\, {r^{-n-1}\over -(n+1)}\Bigg|_a^b
      \={1\over4\pi\varepsilon_0 c(n+1)}\, {b^{n+1}-a^{n+1}\over b^{n+1} a^{n+1}}~.
\ee
\normalsize

\subsection{Cylindrical Capacitor with Dielectric I}

Show that the capacitance of a cylindrical capacitor which is filled with
a dielectric having dielectric constant $\kappa(r)=c r^n$, where $r$
is the distance from the axis and $c$, $n\ne 0$ are constants, is given
by
$$
    C \= 2\pi\varepsilon_0 \, hcn\, { a^n b^n\over b^n-a^n}~.
$$

\footnotesize {\bfseries Solution} 

We partition the capacitor into coaxial infinitesimal cylindrical shells of
radius $r$ and thickness $dr$ which are connected in series
(see figure \ref{fig:hollowcylinder}). 
The infinitesimal incapacitance
of such a shell is
$$
   dD\={1\over\varepsilon_0\kappa(r)}\,{dr\over 2\pi r h}
     \={1\over2\pi\varepsilon_0 hc}\,{dr\over r^{n+1}}~.
$$ 
Therefore
\bb
   D\={1\over2\pi\varepsilon_0 hc}\,\int_a^b {dr\over r^{n+1}}
    \={1\over2\pi\varepsilon_0 hc}\,{b^n-a^n\over n\, a^n b^n}~,
\ee
since $n\ne0$. Therefore
$$
  C\={1\over D} \= 2\pi\varepsilon_0 \, hcn\, { a^n b^n\over b^n-a^n}~,
$$
i.e. exactly the advertised formula.

\normalsize

\subsection{Cylindrical Capacitor with Dielectric II}

Show that the capacitance of a cylindrical capacitor which is filled with
a dielectric having dielectric constant $\kappa(z)=c z^n$, where $z$
is the distance from the base and $c$, $n\ge 0$ are constants, is given
by
$$
    C \= 2\pi\varepsilon_0 \, {ch^{n+1}\over(n+1)\,\ln(b/a)}~.
$$

\footnotesize {\bfseries Solution}

We partition the capacitor into coaxial infinitesimal annuli of
thickness $dz$ which are connected in parallel 
(see figure \ref{fig:hollowcylinder}).
Each infinitesimal capacacitor has the geometry of a cylindrical
capacitor and therefore
its infinitesimal capacitance is given by
$$
   dC\={2\pi\varepsilon_0\kappa(z)}\,{dz\over\ln(b/a)}
     \={2\pi\varepsilon_0 c \over\ln(b/a)} \, z^n \, dz~.
$$
Therefore
\bb
     C \={2\pi\varepsilon_0 c \over\ln(b/a)} \, \int_0^h z^n \, dz
       \={2\pi\varepsilon_0 c \over\ln(b/a)} \, {h^{n+1}\over n+1}~,
\ee
since $n\ge0$.
\normalsize

\subsection{Truncated-Cone Capacitor I}
\begin{figure}[h]
\begin{center}
\psfrag{h}{$h$}
\psfrag{r}{$r$}
\psfrag{z}{$z$}
\psfrag{c}{$c$}
\psfrag{r-b}{$r-b$}
\psfrag{c-b}{$c-b$}
\psfrag{dr}{$dr$}
\psfrag{dz}{$dz$}
\psfrag{a}{$a$}
\includegraphics[width=8cm]{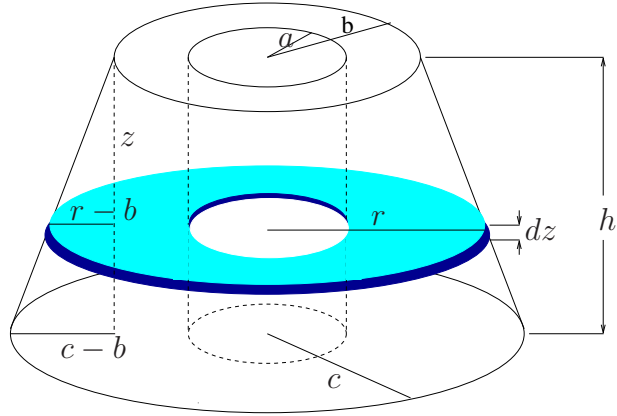}
\end{center}
\caption{A partition of the truncated cone in infinitesimal slices.}
\label{fig:hollowcone}
\end{figure}

(a) Two metallic flat annuli are placed such that they form a
capacitor with the shape of a hollow truncated cone as seen
in figure \ref{fig:hollowcone}. Partition the capacitor into 
infinitesimal capacitors and show that the capacitance 
is given by
$$
    C \= 2\pi\varepsilon\, {h\over a(c-b)}\,
         \lb \ln{c-a\over c+a}-\ln{b-a\over b+a}\rb~.
$$
Show that this result reduces to that of a cylindrical capacitor for
$c=b$. Also, show that it agrees with the result  
for a parallel-plate capacitor with $a=0$.
   
(b) Now, fill the two bases with disks of radius $a$ and argue that
the capacitance of the hollow truncated cone equals that of the
truncated cone minus the capacitance of the parallel-plate
capacitor that we have removed. This means that the capacitance of
the hollow truncated cone should equal 
$$
   C\= \pi\varepsilon_0\, {bc-a^2\over h}~.
$$
How is it possible that this result does not agree with that of part
(a)?

\footnotesize {\bfseries Solution}

(a) We divide the truncated cone into annuli of height $dz$. These are parallel-plate
capacitors connected in series. Therefore
$$
   dD \={1\over\varepsilon_0}\, {dz\over\pi(r^2-a^2)}~.
$$
From the similar triangles see on the left side of figure \ref{fig:hollowcone},
we see that
\beq
 {z\over h}\={r-b\over c-b} ~\Rightarrow~ dz\={h\over c-b}\, dr~.
\label{eq:ratio}
\eeq
Therefore
\bb
   dD \= {1\over\varepsilon_0\pi} \,{h\over c-b}\, {dr\over r^2-a^2} 
      \= {h\over 2\pi\varepsilon_0a(c-b)}\, \lp{1\over r-a}-{1\over r+a}\rp\, dr~,
\ee
and
\bb
   D&=&{h\over 2\pi\varepsilon_0a(c-b)}\,\int_b^c \lp{1\over r-a}-{1\over r+a}\rp\, dr
    \={h\over 2\pi\varepsilon_0a(c-b)}\,\lp\ln{c-a\over c+a}-\ln{b-a\over b+a}\rp~.
\ee 

When $c=b$ we see that $D=0/0$ and therefore the result cannot be found by simple
substitution. However we can use L' Hospital's rule\footnote{L' Hospital's rule
states that: if $\lim_{x\to x_0} f(x)=\lim_{x\to x_0} g(x)= 0$ and
the limit $\lim_{x\to x_0} f'(x)/g'(x)$ exists, then 
$\lim_{x\to x_0} f(x)/g(x)= \lim_{x\to x_0} f'(x)/g'(x)$.}:
\bb
  D\= {h\over 2\pi\varepsilon_0a}\,\lim_{c\to b}{\ln{c-a\over c+a}-\ln{b-a\over b+a}
      \over c-b} 
   \= {h\over 2\pi\varepsilon_0a}\,\lim_{c\to b} {d\over dc} \ln{c-a\over c+a}
   \= {h\over 2\pi\varepsilon_0a}\, {2a\over b^2-a^2}~.
   \= {h\over \pi\varepsilon_0}\, {1\over b^2-a^2}~.
\ee
This is just $C=\varepsilon_0\, {A/h}$ for a plane capacitor.

The case $a=0$ is obtained in the same way:
\bb
  D&=& {h\over 2\pi\varepsilon_0(c-b)}\,\lim_{a\to0}
      {\ln{c-a\over c+a}-\ln{b-a\over b+a}\over a}
   \= {h\over 2\pi\varepsilon_0(c-b)}\,
      \lim_{a\to0}{d\over da} \lp \ln{c-a\over c+a}-\ln{b-a\over b+a}\rp\\
   &=& {h\over 2\pi\varepsilon_0(c-b)}\,
      \lim_{a\to0} \lp {-2c\over c^2-a^2}-{-2b\over b^2-a^2}\rp
   \= {h\over \pi\varepsilon_0 cb}~, 
\ee
as found previously.

(b) We use a parallel-plate capacitor with circular plates of radius $a$
at a distance $h$ to fill the plates of our capacitor. This capacitor
has capacitance 
$$
    C_{add}=\varepsilon_0 \, {\pi a^2\over h}~.
$$
The conical capacitor we have thus created has capacitance 
$$
   C_{total} \= \varepsilon_0 \, {\pi cb \over h}~.
$$
The original capacitor and the one we added are connected in parallel
since the same voltage is applied at their plates. Therefore, according
to the superposition principle
$$
   C_{total} \= C + C_{add} ~\Rightarrow~ C\= \varepsilon_0 \, {\pi(cb-a^2)\over h}~.
$$
Apparently this result does not agree with that of part (a). The reason is
subtle but easy to find. The superposition principle states that
\textit{if a problem in electricity can be split in two other problems, then
the solution to the original problem is the sum of the solutions of
the partial problems}. But is our problem the \textit{exact} sum of the
two partial ones?

Let's assume that each plate of the truncated cone has a charge
of absolute value $Q$ and constant charge density equal to
$\sigma=Q/\pi b^2$ on the top plate and equal to $\sigma'=-Q/\pi c^2$ on
the bottom plate.   

$Q$ splits into $Q_1$ and $Q_2$ on the plates 
of the hollow truncated cone and the cylinder, respectively,
in proportion to the areas of the plates.

On the top plate of the hollow truncated cone we have
$Q_1=\sigma\, \pi (b^2-a^2)$ and on the top plate of the cylinder 
$Q_2=\sigma\,\pi  a^2$.

On the bottom plate of the hollow truncated cone we have
$Q'_1=\sigma' \, \pi (c^2-a^2)$ and on the top plate of the cylinder 
$Q'_2=\sigma'\,\pi  a^2$. However, $Q'_1$ and $Q'_2$ are not $-Q_1$
and $-Q_2$ (except when $b=c$). The only way to ensure this is to change the charge
densities on the plates. But then the problem is \textit{not} a simple
addition of two other problems.

\normalsize

\begin{figure}[h]
\begin{center}
\psfrag{r}{$r$}
\psfrag{dr}{$dr$}
\psfrag{a}{$a$}
\psfrag{b}{$b$}
\psfrag{c}{$c$}
\psfrag{+}{$+$}
\psfrag{=}{$=$}
\includegraphics[width=14cm]{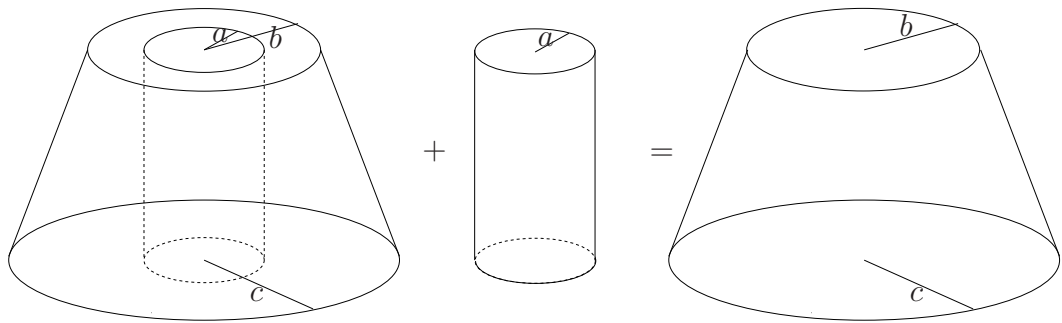}
\end{center}
\caption{The truncated cone is the sum of the hollow truncated cone
         plus a cylinder.}
\label{fig:superposition}
\end{figure}

\subsection{Truncated-Cone Capacitor II}
A capacitor with the shape of a hollow truncated cone is now formed
from two `cylindrical' shells. Find its  capacitance $C$. 

\footnotesize {\bfseries Solution}

As in the previous problem, we partition the capacitor into infinitesimal
annuli. However, in this case they behave as infinitesimal 
\textit{cylindrical} capacitors connected in series. The plates have radii $a$ and $r$,
and the height is $dz$. Therefore
\bb
    dC \=2\pi\varepsilon_0\,{dz\over\ln{r\over a}}~.
\ee
Using equation \calle{eq:ratio}, we can find
\bb
  C \= 2\pi\varepsilon_0\,{h\over c-b} \, \int_b^c {dr\over\ln{r\over a}}~.
    \= 2\pi\varepsilon_0\,{ha\over c-b} \, \int_{b/a}^{c/a} {dt\over\ln t}~,
\ee
where in the last equation we have made a change of variables $t=r/a$.

The integral
$$
   \mbox{li}(x) \eq \int_0^x {dt\over\ln t}
$$
is known as the {\bfseries logarithmic integral} \cite{AS}. Using this definition,
we can write the previous result in the form:
$$
   C \=  2\pi\varepsilon_0\,{ha\over c-b} \, \lb \mbox{li}(c/a)-\mbox{li}(b/a)\rb~.
$$

When $b=c$, we find  $C=0/0$ and therefore we should use L' Hospital's rule
to compute the result:
\bb
   C \=  2\pi\varepsilon_0\,ha\, \lim_{c\to b} {d\over dc} \int_0^{c/a} {dt\over\ln t}~.\ee
Recall now that
\bb
    {d\over dX} \int_0^X  f(t) dt \=  f(X)~.
\ee
Therefore, 
\bb
   C \=  2\pi\varepsilon_0\,ha\, {1\over a\ln(b/a)}
     \=2\pi\varepsilon_0\, {h\over \ln(b/a)}~,
\ee
i.e. the result of a cylindrical capacitor with radii $a$ and $b$.
\normalsize

\subsection{Hollow Cylindrical Conductor}
\label{sec:cyl-conductor}

A conductor has the shape of a hollow cylinder as seen in figure
\ref{fig:hollowcylinder}.
Show that the resistance
when the voltage is applied between the inner and outer surfaces
is given by
$$
   R \={\rho\over 2\pi h}\, \ln{b\over a}~.
$$

\footnotesize {\bfseries Solution}
 
We split the conductor into infinitesimal cylindrical shells (left picture
of figure \ref{fig:hollowcylinder}) which are connected in series and have infinitesimal
resistance:
$$
   dR\=\rho\, {dr\over2\pi r h}~.
$$
The total resistance of the conductor is then:
\bb
   R\=\int dR\= {\rho\over2\pi h} \, \int_b^c {dr\over r}
    \= {\rho\over2\pi h} \, \ln{c\over b}~.
\ee

\normalsize

\subsection{Hollow Truncated-Cone Conductor I}

(a) A conductor has the shape seen in figure \ref{fig:hollowcone}.
Show that the resistance
when the voltage is applied between the upper and lower
bases is given by
$$
  R\=\rho\,{h\over 2\pi a(c-b)}\,
     \lb \ln{c-a\over c+a}-\ln{b-a\over b+a}\rb~.
$$
Show that this result reduces to that of a solid truncated-cone
for $a=0$. 

(b) Argue now that the resistance of the hollow truncated-conical
wire is the difference between the the resistance of the truncated-conical
wire and a cylindrical wire of radius $a$. This implies that 
$$
     R\=\rho\,{bc-a^2\over h}~.
$$
Explain why this does not agree with part (a).

\footnotesize {\bfseries Solution}

(a) We partition the conductor into infinitesimal cylindrical conductors of
length $dz$ and cross-sectional area $\pi(r^2-a^2)$ as seen in figure
\ref{fig:hollowcone}. These infinitesimal conductors are connected in series.
They have resistance
$$
    dR \= \rho\, {dz\over \pi(r^2-a^2)}
       \= \rho\,{h\over \pi(c-b)}\, {dr\over r^2-a^2}~,
$$
where we used equation \calle{eq:ratio}.
Therefore
\bb
   R\=\rho\,{h\over\pi(c-b)}\,\int_b^c {dr\over \pi(r^2-a^2)}
    \=\rho\,{h\over2\pi(c-b)a}\,\lb\ln{c-a\over c+a}-\ln{b-a\over b+a}\rb~.
\ee

For $a=0$, $R=0/0$. We thus must use L' Hospital's rule:
\bb
    R&=&\rho\,{h\over2\pi(c-b)}\,\lim_{a\to0}
        {\ln{c-a\over c+a}-\ln{b-a\over b+a}\over a}
     \=\rho\,{h\over2\pi(c-b)}\,\lim_{a\to0}{d\over da}
        \lb\ln{c-a\over c+a}-\ln{b-a\over b+a}\rb
     \=\rho\,{h\over\pi cb}~.
\ee

(b) Using the superposition principle for figure \ref{fig:superposition},
    we would have written down
$$
    D\=D_{cone}-D_{cylinder}\={1\over\rho}\,{cb-a^2\over h}~,
$$
since they are connected in parallel. Then
$$
    R\=\rho\,{h\over\pi (cb-a^2)}~.
$$
This does not agree with the result of part (a) since the sum of two partial
problems is \textit{not} the problem we are studying. One can verify this
by checking the current densities on the top and bottom faces.
\normalsize

\subsection{Hollow Truncated-Cone Conductor II}

A conductor has the shape seen in figure \ref{fig:hollowcone}.
Show that the resistance
when the voltage is applied between the inner and outer surfaces
is given by
$$
   R \={\rho\over 2\pi h a}\, {c-b\over{\rm li}(c/a)-{\rm li}(b/a)}~.
$$
Show that, for $c=b$, this result agrees with that of
problem \ref{sec:cyl-conductor}.

\footnotesize {\bfseries Solution}

We partition the conductor into infinitesimal annuli connected in parallel.
The corresponding infinitesimal conductivity is 
$$
   dS\=2\pi\sigma\, {dz\over\ln{r\over a}}
     \=2\pi\sigma\,{h\over c-b}\, {dr\over\ln{r\over a}}~.
$$
From this, the total conductivity is found to be
\bb
   S\=2\pi\sigma  \, \int_b^c {dz\over\ln{r\over a}}
    \=2\pi\sigma\,{ha\over c-b}  \, \lb\mbox{li}(c/a)-\mbox{li}(b/a)\rb~,
\ee
which can be inverted to give to total resistance
\bb
    R\={\rho\over2\pi ha}  \, {c-b\over\mbox{li}(c/a)-\mbox{li}(b/a)}~.
\ee

\normalsize

\subsection{Wedge Conductor}
\label{sec:wedge}

\begin{figure}[htb!]
\begin{center}
\psfrag{x}{$x$}
\psfrag{y}{$y$}
\psfrag{dy}{$dy$}
\psfrag{z}{$z$}
\psfrag{a}{$a$}
\psfrag{b}{$b$}
\psfrag{c}{$c$}
\psfrag{c-b}{$c-b$}
\psfrag{z-b}{$z-b$}
\psfrag{l}{$\ell$}
\includegraphics[width=14cm]{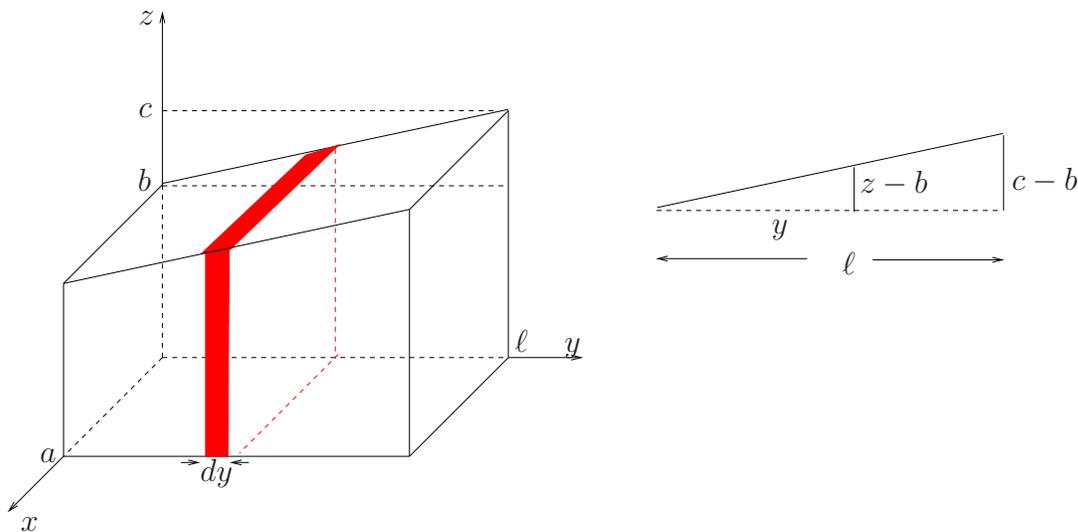}
\end{center}
\caption{A wedge partitioned in infinitesimal layers.} 
\label{fig:wedge}
\end{figure}

A conductor has the shape of a truncated wedge as seen in figure
\ref{fig:wedge}.
Show that the resistance of the conductor
when the voltage is applied between the left and right faces
is 
$$
   R\= {\rho\over a}\, {\ell \,\ln(c/b)\over c-b}~,
$$
while the resistance when the voltage is applied between the top and
bottom faces is
$$
   R\= {\rho\over a}\, {c-b\over\ell\,\ln(c/b)}~.
$$

\footnotesize {\bfseries Solution}

(a) We partition the conductor into infinitesimal layers of thickness $dy$ 
    along the $y$-direction. These layers are infinitesimal resistors
    with the shape of square cylinders connected in series. Their resistance is
$$
   dR\=\rho\,{dy\over z a}~.
$$
From the figure we see that
$$
  {y\over \ell}\={z-b\over c-b}~\Rightarrow~
   y\={\ell\over c-b}\, (z-b) ~\Rightarrow~
   dy\={\ell\over c-b}\, dz.
$$
Then
\bb
  dR\=\rho\, {\ell\over a(c-b)}\, {dz\over z}~,
\ee
and
\bb
  R\=\rho\, {\ell\over a(c-b)}\,\int_b^c {dz\over z}
   \=\rho\, {\ell\over a(c-b)}\,\ln{c/b}~.
\ee

(b) When the voltage is applied between the top and bottom faces, the
infinitesimal resistors are connected in parallel. Now the current flows
through area $a\, dy$ and the length it travels is $z$:
$$
    dS\=\sigma\, {ady\over z}
      \=\sigma\, {a\ell\over c-b} \, {dz\over z}~.
$$
Therefore
\bb
    S\= \sigma\, {a\ell\over c-b} \,\int_b^c {dz\over z}
     \= \sigma\, {a\ell\over c-b} \,\ln{c\over b}~,
\ee
from which
\bb
   R\= \rho\,{c-b\over a\ell}\, {1\over \ln(c/b)}~.
\ee 
\normalsize

\subsection{Toroidal Inductor}
\label{sec:toroid}

Find the inductance for a section of angular span $\phi$ of a toroidal inductor of 
radii $a$ and $b$, height $h$, and $N$ number of turns. 

\footnotesize {\bfseries Solution}

\begin{figure}[htb!]
\begin{center}
\psfrag{I}{$I$}
\psfrag{a}{$a$}
\psfrag{b}{$b$}
\psfrag{r}{$r$}
\psfrag{dr}{$dr$}
\psfrag{h}{$h$}
\psfrag{s}{$s$}
\psfrag{f}{$\phi$}
\includegraphics[width=11cm]{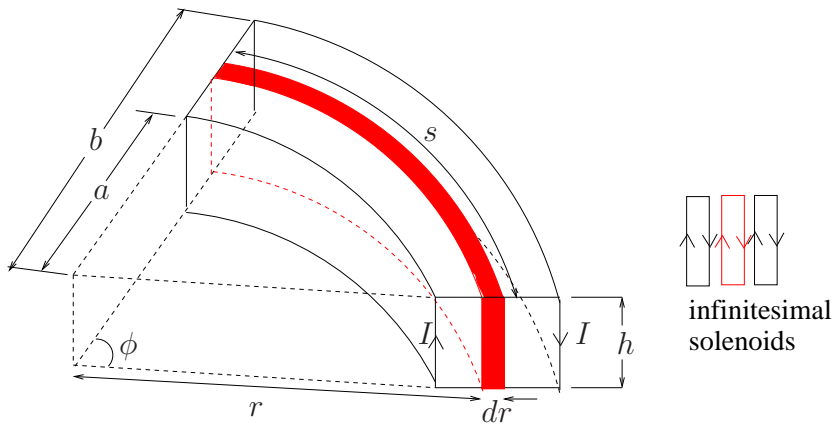}
\end{center}
\caption{A toroidal inductor.}
\label{fig:toroid}
\end{figure}

We partition the toroid into infinitesimal solenoids all run by the same current
and thus connected in series.
This is seen in figure \ref{fig:toroid}.

All infinitesimal inductors
have the same number of turns $N$ and they have the simple geometry of
a solenoid. Therefore, each has an infinitesimal inductance 
$$ 
   dL\=\mu_0\, N^2\,{h\,dr\over s}~,
$$
where $s=\phi r$ is the length of the inductor at distance $r$ from the center.
Then
\bb
   L \= \int dL &=& \mu_0\, N^2\,{h\over \phi}\, \int_a^b {dr\over r}
     \= \mu_0\, N^2\,{h\over \phi}\, \ln{b\over a}~.
\ee
For a full circle, $\phi=2\pi$ and
$$
   L\=\mu_0\, N^2\,{h\over 2\pi}\, \ln{b\over a}~,
$$
 a well-known result \cite{HRW,Tipler,WP}, usually found by computing flux. 
Notice that, using our method, it is not necessary to know the value of the magnetic
field in order to find the inductance.
\normalsize

\subsection{Parallel-Plate Inductor}
\label{sec:parallel-ind}

Split the parallel-plate inductor into convenient infinitesimal inductors.
Then make use of equation
\calle{eq:Ldef} to again derive  equation \calle{eq:Lbasic}.

\footnotesize {\bfseries Solution}

\begin{figure}[htb!]
\begin{center}
\psfrag{dx}{$dx$}
\psfrag{d}{$d$}
\psfrag{w}{$w$}
\psfrag{L}{$L$}
\psfrag{I}{$I$}
\includegraphics[width=8cm]{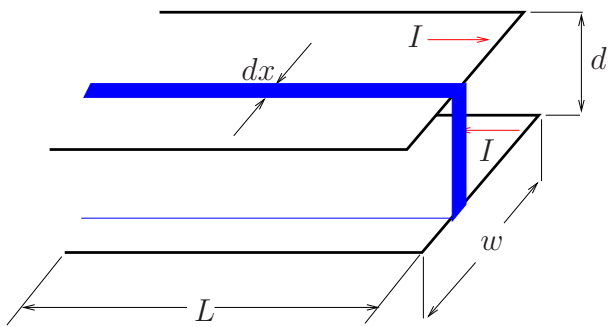}
\end{center}
\caption{A parallel-plate inductor can be split into infinitesimal inductors
         connected in parallel.}
\label{fig:plane-inductor2}
\end{figure}

The inductor is split in parallel infinitesimal slices as seen
in figure \ref{fig:plane-inductor2}. Each slice is similar to a turn of
a solenoid; it is carrying an infinitesimal current
$dI=J_s dx$. The infinitesimal slices have a deductance of
$$
   dK\= {dI\over\Phi_B}~,
$$ 
where $\Phi_B=B L d=\mu_0 J_s L d$. Therefore
$$
   dK\=  {1 \over \mu_0  L d}\, dx~.
$$
The total deductance is then
$$
   K\=\int_0^w {1 \over \mu_0  L d}\, dx \= {w \over \mu_0  L d}~,
$$
and
$$
   L\={1\over K}\= \mu_0\, {L\, d\over w}~.
$$

\normalsize

\subsection{Coaxial Inductor I}
\label{sec:coaxial}

Think of a coaxial cable made of two cylindrical shells of radii $a$ and $b$.
If the cable has length $h$, compute its inductance. (The currents are uniformly
distributed along the cross-sections of the wires.)

\footnotesize {\bfseries Solution}

\begin{figure}[ht!]
\begin{center}
\psfrag{r}{$r$}
\psfrag{dr}{$dr$}
\psfrag{a}{$a$}
\psfrag{b}{$b$}
\psfrag{h}{$h$}
\includegraphics[height=9cm]{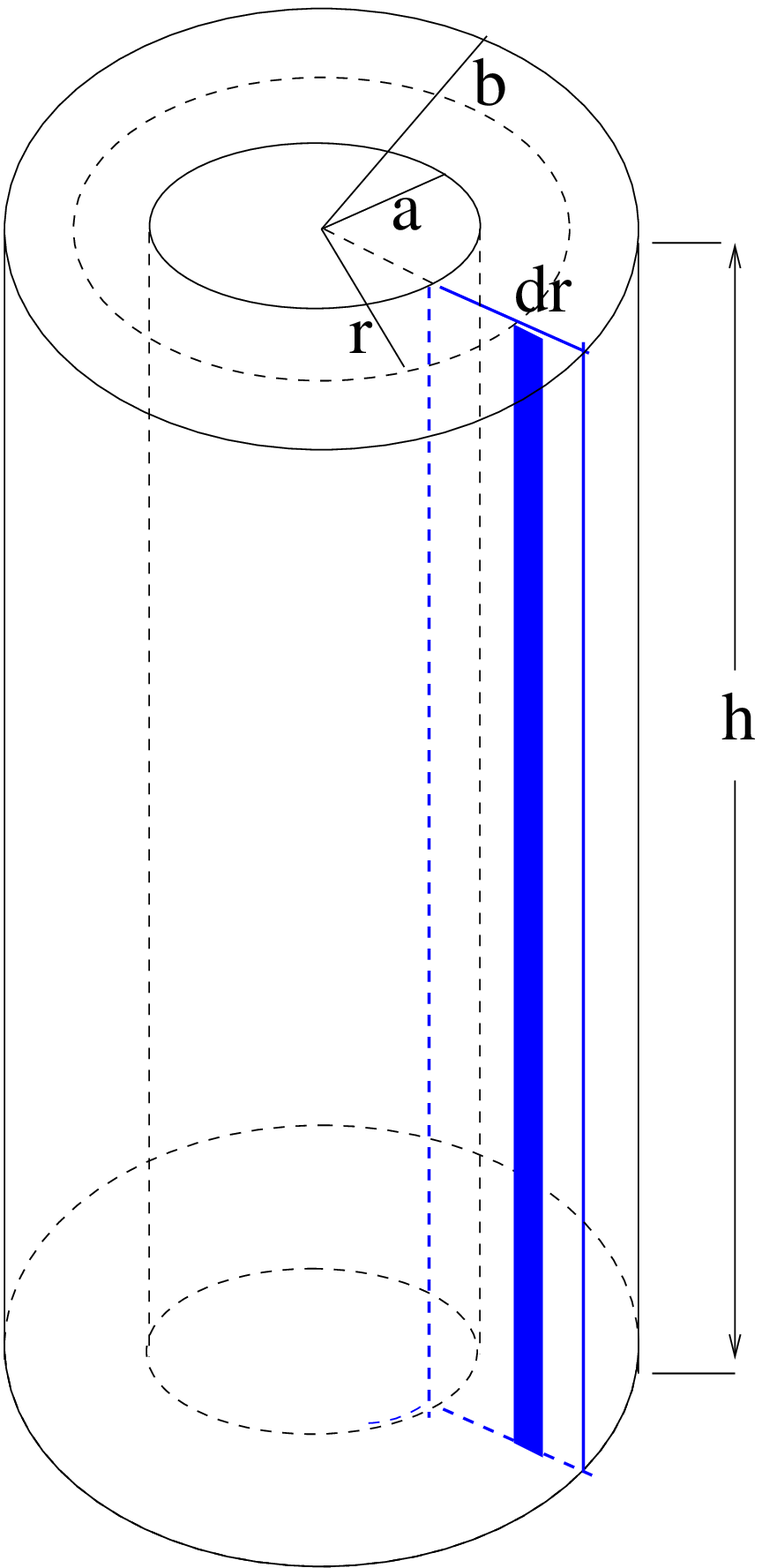}
\hspace{10mm}
\psfrag{r}{$r$}
\psfrag{dr}{$dr$}
\psfrag{a}{$a$}
\psfrag{b}{$b$}
\psfrag{h}{$h$}
\psfrag{df}{$d\phi$}
\includegraphics[height=9cm]{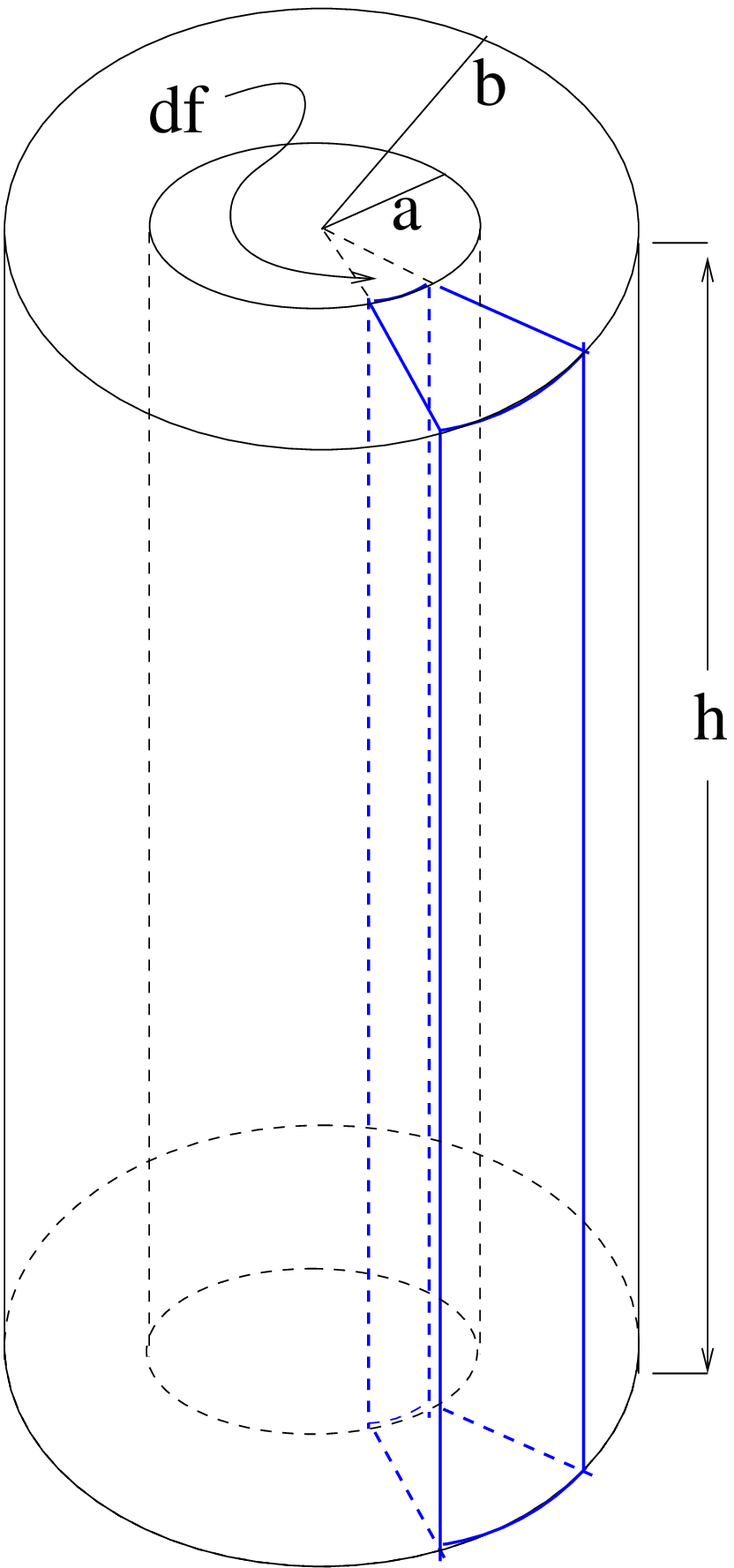}
\hspace{1cm}
\psfrag{r}{$r$}
\psfrag{dr}{$dr$}
\psfrag{a}{$a$}
\psfrag{b}{$b$}
\psfrag{h}{$h$}
\includegraphics[height=10cm]{cylinders.eps}
\end{center}
\caption{A coaxial cable made from two cylindrical shells with radii $a$ and
         $b$ and height $h$.
         In the right picture, we have sliced it into infinitesimal cylindrical
         shells, while in the middle picture we have sliced it into
         infinitesimal  wedges.}
\label{fig:coaxial}
\end{figure}

\underline{Standard Calculation}:
The standard computation found in introductory physics books (e.g. \cite{WP}) uses
the definition \calle{eq:Ldef}. The magnetic field between the two cylinders
can be easily found using Amp\`ere's law:
$$
   B\={\mu_0 I\over 2\pi r}~,
$$
where $r$ is the distance from the center. Then we take a cross-section
between the two cylinders (seen in blue in the left picture of figure 
\ref{fig:coaxial}). We split this cross-section into infinitesimal strips
of width $dr$ that are penetrated by infinitesimal flux 
$$
   d\Phi_B\=B\,h\,dr\= {\mu_0 I h\over 2\pi }\, {dr\over r}~.
$$
The total flux through the whole cross-section is
$$
   \Phi_B\=\int d\Phi_B\= {\mu_0 I h\over 2\pi }\, \int_a^b {dr\over r}
         \= {\mu_0 I h\over 2\pi }\,\ln{b\over a}~.
$$
Then, $L=\Phi_B/I$ and therefore
$$
   L\= {\mu_0  h\over 2\pi }\,\ln{b\over a}~.
$$

Obviously the standard computation falls short of explaining why only one cross-section
has been used and no integration over the angular coordinate has been 
performed.

\underline{New Calculation I}:
 Let's partition the cable into infinitesimal inductors with the shape
of the wedge as seen in the middle picture of figure \ref{fig:coaxial}.
The angular span of each wedge is $d\phi$ and each plate carries current
$dI=I\,{d\phi\over 2\pi}$. The wedges are connected in parallel.

Each wedge has an infinitesimal deductance of
$$
   dK\= {dI\over \Phi_B}\={I\over2\pi\Phi_B}\, d\phi~.
$$
The total detuctance is
$$
   K\={I\over2\pi\Phi_B}\,\int_0^{2\pi} d\phi\={I\over\Phi_B}~.
$$
and therefore we recover the result of the previous solution.
We now see why only the flux of a single cross-section should
be included. 

\underline{New Calculation II}:
The computation we presented in the previous paragraph was no  shorter
than the standard one, as it also must compute the flux for one 
cross-section. It only had the advantage of explaining why the flux of a single
cross-section must be included. However, now we give a short calculation
based on our technique, by-passing the use of flux and relying solely on
the basic formula \calle{eq:Lbasic}.

We partition the cable into infinitesimal cylindrical shells seen in the right
picture of figure \ref{fig:coaxial} which behave as
parallel-plate inductors connected in series. The distance
between the plates is $dr$, the length of each plate is $h$, and the
width of each plate is $2\pi r$. Therefore
$$
    dL\= \mu_0\,{h \, dr\over 2\pi r}~.
$$
The total inductance is then
$$
    L\= \mu_0\,{h\over2\pi}\,\int_a^b {dr\over r}\=\mu_0\,{h\over2\pi}\,\ln{b\over a}~.
$$
\normalsize

\subsection{Coaxial Inductor II}
\label{sec:coaxial2}

In the coaxial inductor of the previous problem,
imagine that the current is flowing from the inner cylindrical shell towards
the outer shell. 
Compute the inductance of this configuration of this configuration.

\footnotesize {\bfseries Solution}

This is another problem that presents the power of our approach. Using
the standard computations, students would have great difficulty in
solving this  problem, as they must first compute the magnetic field. However,
our method reduces the problem to a an almost trivial calculation!

We again partition the cable into cylindrical shells of width $dr$.
These are now infinitesimal inductors that are connected in parallel,
each having a deductance
$$
   dK\={1\over\mu_0}\,{dr\over2\pi r h}~.
$$
Therefore the total deductance is
\bb
   K\=\int dK\= {1\over\mu_0}\,{1\over2\pi h}\, \int_a^b {dr\over r}
    \= {1\over\mu_0}\,{1\over2\pi h}\, \ln{b\over a}~,
\ee
which of course implies that
$$
   L\= \mu_0\, 2\pi\, {h\over\ln{b\over a}}~.
$$
\normalsize

\section{Conclusions}

There is probably no need for additional problems. The reader has certainly
uncovered the pattern. All the quantities we have used
---let the generic symbol $P$ stand for any of
them---follow a simple additive law
\bb
   P &=& \sum_i P_i~,~~~~~\mbox{discrete~case}~,\\ 
   P &=& \int dP~,~~~~~\mbox{continuous~case}~.
\ee 
For elements in the shape of a uniform cylinder of length $L$
and cross-section $A$
whose material is characterised by the constant $p$ (corresponding to
quantity $P$), $P$ given by
$$
   P \=
   \cases{ \displaystyle 
           p\, {L\over A}~, & if~$P$~stands~for~$R,D,L,\ell,\CR$~,\cr
            \cr
           \displaystyle
           {1\over p}\,{A\over L}~, & if~$P$~stands~for~$G,C,K,k,\CG$~.\cr
         }
$$
For infinitesimal elements
$$
   dP \=
   \cases{ \displaystyle
           p\, {dL\over A}~, & if~$P$~stands~for~$R,D,L,\ell,\CR$~,\cr
            \cr
           \displaystyle
           {1\over p}\,{dA\over L}~, & if~$P$~stands~for~$G,C,K,k,\CG$~.\cr
         }
$$

In all cases of identical geometry, the results will be identical.
In fact, in many instances above we could have saved some computations
but we avoided doing so in order to present the big picture first. Once the
reader is aware of the global picture, she can easily use it to transfer
a result for a quantity $P$ in some geometry to a quantity
$P'$ in a similar geometry. An example of this is as follows.
\begin{center}
\begin{minipage}{11cm}
 {\footnotesize 
  A sauna room has the shape of the wedge seen in figure \ref{fig:wedge}.
  The insulation of the wall at $y=\ell$ has deteriorated and,
  as a result, the wall is at a lower temperature with respect to the
  wall at $y=0$. An engineer would like to know the thermal resistance of 
  the room. Assume that all other walls are at the same temprature with
  the wall at $y=0$.
 }
\end{minipage}
\end{center}
It is obvious that heat will flow from the left wall to the right wall.
One might want to divide the room in infinitesimal layers parrallel to the side
walls and continue as usual but this is not necessary. This problem has already
been solved in section \ref{sec:wedge}. Based on the results there, we can write
dowm immediately that
$$
  \CR_{th}\={\rho_{th}\over a}\, {\ell\,\ln(c/b)\over c-b}~.
$$
Here is another problem
\begin{center}
\begin{minipage}{11cm}
 {\footnotesize
  A cylindrical cable of radius $R$ and lenght $h$
  is made of a huge number of small filaments
  such that the cable may be considered continuous. The filaments
  have been arranged in such a way that Young's modulus for the
  cable is $Y(r)=c\,r$, where $r$ is the distance from the center of the
  cable and $c$ some constant. What is the stiffness of the cable?
 }
\end{minipage}
\end{center}
The stiffeness of a cylindrical shell of radius $r$ is given by
$$
   dk\= Y(r)\,{2\pi rdr\over h}~.
$$
The stiffness  of the cable would then be
$$
   k\= {2\pi c\,R^3\over 3h}~.
$$

The reader is invited to construct similar problems for quantities in
table \ref{table:2}.



\begin{thebibliography}{9}
\footnotesize

\bibitem{AS}
  \textsc{M. Abramowitz, I.A. Stegun},
  \textsf{Handbook of Mathematical Functions with Formulas},
   Dover.
\bibitem{Alonso}
  \textsc{M. Alonso, E.J. Finn},
  \textsf{Physics},
  Addison-Wesley.
\bibitem{HRW}
  \textsc{D. Haliday, R. Resnick, J. Walker},
  \textsf{Fundamentals of Physics}, 6th ed.,
   John-Wiley \& Sons.
\bibitem{Hecht}
  \textsc{E. Hecht},
  \textsf{Physics},
  Brooks/Cole 1996.
\bibitem{Nolan}
  \textsc{P. Nolan},
  \textsf{Fundamentals of College Physics},
  Wm. C. Brown Communications 1993.
\bibitem{Romano}
  \textsc{J.D. Romano, R.H. Price},
  \textit{The Conical Resistor Conundrum: A Potential Solution},
  Am. J. Phys. {\bfseries 64} (1996) 1150.
\bibitem{Serway}
  \textsc{R.A. Serway},
  \textsf{Physics for Scientists and Engineers}, 4th ed.,
  Saunders College Publishing.
\bibitem{Tipler}
  \textsc{P.A. Tipler},
  \textsf{Physics for Scientists and Engineers}, 3rd ed.,
  Worth Publishers.
\bibitem{WP}
  \textsc{R. Wolfson, J.M. Pasachoff},
  \textsf{Physics for Scientists and Engineers}, 3rd ed.,
  Addison-Wesley.
\bibitem{YF}
  \textsc{H.D. Young, R.A. Freedman},
  \textsf{University Physics}, 9th ed.,
  Addison-Wesley 1996.
\end{thebibliography}
\end{document}